\documentclass[twocolumn,showpacs,preprintnumbers,amsmath,amssymb]{revtex4}

\usepackage{graphicx}
\usepackage{dcolumn}
\usepackage{bm}

\begin{document}

\title{Population control of 2s-2p transitions in hydrogen}

\author{Kh.Kh.Shakov}
\author{J.H.McGuire}
\affiliation{Physics Department, Tulane University, New Orleans, LA 70118 USA.}

\date{\today}

\begin{abstract}
We consider the time evolution of the occupation probabilities for the $2s-2p$ transition in a hydrogen atom interacting with an external field, $V(t)$. 
A two-state model and a dipole approximation are used.
In the case of degenerate energy levels an analytical solution of the time-dependent Shr\"odinger equation for the probability amplitudes exists.
The form of the solution allows one to choose the ratio of the field amplitude to its frequency that leads to temporal trapping of electrons in specific states.
The analytic solution is valid when the separation of the energy levels is small compared to the energy of the interacting radiation.
\end{abstract}

\pacs{32.80.Qk,42.50.-p}

\maketitle

\section{Introduction}
The ability to control quantum systems, e.g. to cause a complete transfer of population from an initial state to a desired final state using an external field, is crucial in quantum computation \cite{Nielsen,Bouwmeester} and modification of reaction dynamics \cite{Rabitz93,Shapiro02}. 
Since the 1970s many techniques for controlling single quantum systems have been developed, e.g. methods for trapping atoms \cite{Javanainen} and arranging atoms on surfaces \cite{Heller}. 
In this paper we consider transitions between the $2s$ and the $2p$ levels in hydrogen caused by an interaction with a microwave radiation field. 
Hydrogen is an instructive example, since it is the simplest naturally occuring atom, and consequently one of the few that permit detailed and precise comparison with theoretical models. 
If one limits the number of active states to two, one has a qubit. 

Transitions between two discrete states can be, under certain conditions, accurately described within the two-state model \cite{Milonni88, McDowell, Shore, Eberly75, Kyrola, Bialynicka, Fritsch}.
The energy separation between the $2s$ and $2p$ levels, discovered in 1947 by Lamb and Retherford \cite{Lamb}, is small compared to the nearest levels, $1s$ and $3p$.
This enables one to use the two-state model over a wide range of field frequencies.
The interaction between a two-state atom and a single mode radiation field is often treated using the rotating wave approximation (RWA) that makes the problem exactly solvable \cite{Eberly75,Milonni88,Mittleman,Silverman}. 
It was first discussed in detail in 1963 by Jaynes and Cummings \cite{Jaynes63}. 
Although RWA provides an accurate description in many cases of interest, it has its limitations as well. 
It requires the frequency of the radiation field to be close to the transition frequency, and it cannot be applied to a system with nearly degenerate energy levels.

Another analytic solution of the time-dependent Shr\"odinger equation exists in the limit of degenerate energy levels. 
We present this solution here. We show that it has an interesting feature which can be used in the problem of quantum control.

Detailed analysis is given of the applicability of our analytical solution. 
Calculations are presented for $2s-2p$ transitions in atomic hydrogen.

Atomic units are used throughout the paper.

\section{Formalism}
In the two-state approximation the exact time-dependent wave function has the form
\begin{equation}\label{f1}
\Psi (\vec{r},t) = a_1(t) \Phi_1(\vec{r}) + a_2(t) \Phi_2(\vec{r}) \ ,
\end{equation}
where $\Phi_n(\vec{r})$ are the orbital functions, and $a_n$ are the probability amplitudes.
The time variation of the probability amplitudes is determined by the time-dependent Schr\"odinger equation, namely, 
\begin{eqnarray}\label{f2}
&& i \dot{a}_1 = E_1 a_1 + V_{12}a_2 \nonumber\\
&& i \dot{a}_2 = E_2 a_2 + V_{21}a_1 \ ,
\end{eqnarray}
where $V_{ij}=\int \Phi_i^*(\vec{r})V_{ext}(\vec{r},t) \Phi_j(\vec{r}) \ d^3r$, and $E_n$ are the electronic energy levels. The diagonal matrix elements $V_{11}$ and $V_{22}$ are required to be zero by the dipole selection rule.

In the dipole limit, an external electromagnetic field interacts with an electron via the time-dependent potential
\begin{equation}\label{f3}
V_{ext}(\vec{r},t)=-\vec{r}\vec{E}(t) \ ,
\end{equation}
where $\vec{r}$ is a relative electron-nuclear distance, and $\vec{E}$ is an electric field which for the plane monochromatic wave can be written as $\vec{E}=\hat{\epsilon}E_0\cos{(\omega t)}$.
The matrix elements $V_{ij}$ can then be expressed as
\begin{equation}\label{f4}
V_{ij}(t)=-\vec{r}_{ij}\hat{\epsilon}E_0\cos{(\omega t)} \ ,
\end{equation}
where $\vec{r}_{ij}=\vec{r}_{ji}^* =\int \Phi_i^*(\vec{r})\vec{r}\Phi_j(\vec{r}) \ d^3r$. 
We define $\omega_{21}=E_2-E_1$, and introduce the Rabi frequency \cite{Milonni88}
\begin{equation}\label{f5}
\chi=\vec{r}_{21} \hat{\epsilon} E_0 \ .
\end{equation}
The arbitrary zero of energy is set at $E_1$, so $E_2 \rightarrow E_2-E_1=\omega_{21}$.
Then equations for the probability amplitudes become
\begin{eqnarray}\label{f6}
&& i \dot{a}_1 = -\chi \cos (\omega t) a_2  \nonumber\\
&& i \dot{a}_2 =  \omega_{21}a_2 - \chi \cos (\omega t) a_1 \ .
\end{eqnarray}
These equations can be solved numerically for an arbitrary values of energy separation $\omega_{21}$ between electronic energy levels, a field frequency $\omega$, and a Rabi frequency $\chi$ that expresses the field-atom interaction in frequency units. 

Analytical solutions exist in various limits. 
When both $\omega$ and $\omega_{21}$ are large compared to the frequency of oscillation of the transition amplitudes and are close one to another, an analytical solution can be obtained either under RWA or within the oscillating-field theory \cite{Silverman}.
Equations (\ref{f6}) decouple when either $\chi \rightarrow 0$ or $\omega_{21} \rightarrow \infty$.
Replacing an oscillating field by  a stationary field (i.e. taking the limit $\omega \rightarrow 0$) also leads to an analytically solvable problem.
Another analytic solution, considered in detail below, exists in the limit of degenerate energy levels.

\subsection{Analytic solution: limit $\omega_{21} \rightarrow 0$}\label{Sec:AnSol}
Equations (\ref{f6}) can be solved analytically in the limit $\omega_{21} \rightarrow 0$, 
namely \cite{Shore, Kyrola, Bialynicka},
\begin{eqnarray}\label{f7}
&& a_1=  \cos \lbrack  \frac{\chi}{\omega} \sin(\omega t) \rbrack \nonumber\\
&& a_2=i \sin \lbrack  \frac{\chi}{\omega} \sin(\omega t) \rbrack \ .
\end{eqnarray}
Here we take the initial conditions as $a_1(0)=1$ and $a_2(0)=0$ , i.e. initially the system is in state 1.

As we can see from (\ref{f7}), the temporal behavior of the probability amplitudes in this limit is defined by the ratio $\chi / \omega$. The argument of the "outer" function (e.g., $\sin$ for $a_2$) varies between $\pm \chi / \omega$. Thus, the transition amplitude oscillates between $0$ and $\sin(\chi / \omega)$.
Note that if  $\chi / \omega < \pi / 2$, the amplitude $a_2$ (and, consequently, the occupation probability $P_2$) never reaches 1, i.e. complete population of the second level cannot be achieved.
If  $\chi / \omega > \pi / 2$, the amplitudes oscillate in a quasi-regular fashion.
 
\subsection{Complete population transfer}\label{2B}
An interesting solution occurs when the ratio $ \chi / \omega = \pi / 2$. Then Eqs (\ref{f7}) take the form
\begin{eqnarray}\label{f8}
&& a_1=  \cos \lbrack  \frac{\pi}{2} \sin(\omega t) \rbrack \nonumber\\
&& a_2=i \sin \lbrack  \frac{\pi}{2} \sin(\omega t) \rbrack \ .
\end{eqnarray}
The corresponding occupation probabilities $P_1=|a_1|^2$ and $P_2=|a_2|^2$ are
\begin{eqnarray}\label{f9}
&& P_1=  \cos^2 \lbrack  \frac{\pi}{2} \sin(\omega t) \rbrack \nonumber\\
&& P_2=  \sin^2 \lbrack  \frac{\pi}{2} \sin(\omega t) \rbrack \ .
\end{eqnarray}
\begin{figure}
\scalebox{0.7}{\includegraphics{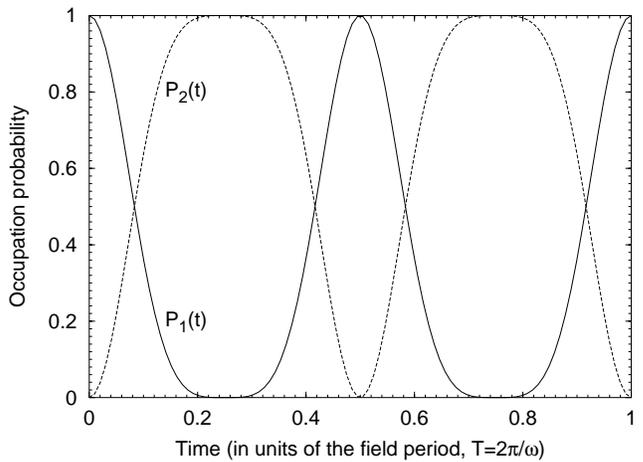}}
\caption{Analytical solution for $\chi/\omega=\pi/2$ for a system with two degenerate energy levels.}
\label{fig1}
\end{figure}
For this special value of $\chi/\omega$ the probability is completely transfered back and forth between the two levels with the period $\pi/\omega$ as shown in figure \ref{fig1}.
It can be easily shown that when $P_2(t)$ reaches its maximum value at $t=\pi/2\omega$, a small deviation $\epsilon$ from the value $\pi/2$ in the ratio $\chi/\omega$ reduces $P_2$ from $1$ to $1-\epsilon^2$. This can be used to estimate how sensitive this approach is to fluctuations in the strength of the external field, $\chi$.
The period of oscillation of the occupation probabilities is two times smaller than the period of oscillation of the radiation field since $P_i=|a_i|^2$.

Complete population transfer occurs in general wherever the action, $A(t)=\int_0^t V_{21}(\tau) \ d \tau$, is an integer multiple of $h/4$.
Each cycle has a "flat" part where the probability value remains very close to the extreme values of 0 or 1 for an extended period of time (cf. figure \ref{fig1}).
As we can see, the external radiation field with the ratio $\chi / \omega=\pi / 2$ can be used to cause periodical population inversion of the electronic levels (a physical realization of a quantum controllable system).
An analysis of the applicability of this analytic solution to $2s-2p$ transitions in hydrogen, including the criteria for the choice of a field frequency, is given in Section \ref{Sec:Res}.
Modifying the shape of $V(t)$ to further control population transfer is also discussed.
 
\subsection{Duration of the population transfer}
Equations (\ref{f9}) can be expanded in a power series in the vicinity of the point 
$t_0=T/4$, where $P_2(t)$ reaches its first maximum. Then $P_2$ can be accurately
approximated by taking only first few terms in the series, which converges rapidly 
for $\tau < T =2 \pi/\omega$. For the form of the potential chosen in Eq.(\ref{f3}), 
the first, second and third derivatives of the function $P_2$ are zero at $t=t_0$, and
\begin{equation}\label{f12}
P_2(t_0+\tau)= 1+\frac{1}{4!} \frac{d^4 P_2}{d t^4}\tau^4 + {\cal O}(\tau^6)
\approx 1 - \frac{\pi^2}{16}(\omega \tau)^4 \ .
\end{equation}
Hence the occupation probability $P_2$ in the vicinity of its maximum can be approximated by a $4th$ degree polynomial. 

\begin{figure}
\scalebox{0.7}{\includegraphics{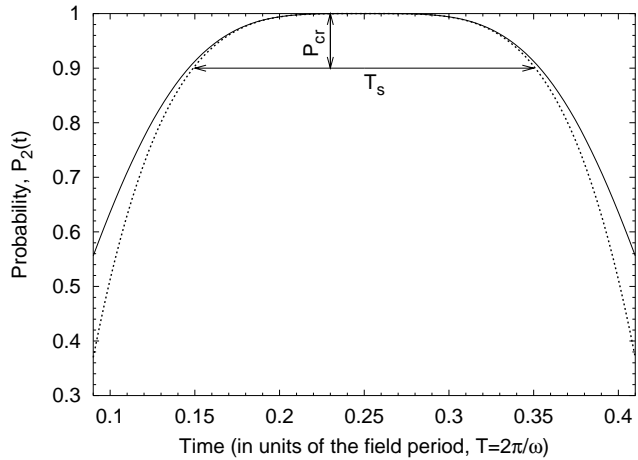}}
\caption{Population leakage in the vicinity of $t_0=T/4$. Solid line - analytic solution (\ref{f9}), dash-dotted line - polynomial approximation (\ref{f12}).} 
\label{fig2}
\end{figure}

The polynomial approximation of Eq.(\ref{f12}) tells one how to choose the field frequency to provide a desired duration of the populated state. To obtain a duration $T_s$ of the populated state with a population leakage less than some critical value $P_{cr}$,
the field frequency to be used is
\begin{equation}\label{estim1}
\omega = \sqrt{\frac{4}{\pi}}\frac{(P_{cr})^{\frac{1}{4}}}{T_s}
\end{equation}
Near $\tau =0$ the error is ${\cal O}((\omega\tau)^6)$.
This is illustrated in figure \ref{fig2}.

\section{Results}\label{Sec:Res}
In this section we compare full and analytic results for the $2s-2p$ transition in hydrogen in the two state approximation.

\subsection{Population leakage}

To obtain our analytic solution to Eq.(\ref{f6}), we had to neglect the $\omega_{21}a_2$ term in the equation 
$ i \dot{a}_2 =  \omega_{21}a_2 - \chi \cos (\omega t) a_1. $
This approximation requires that $\omega_{21} \ll \chi$, or (since  $\chi/\omega=\pi/2$), $\omega_{21} \ll \omega.$
When $\omega_{21}$ is finite, the population transfer is not complete.
The effect of this population leakage can be calculated by expanding the transition amplitude $a_2$ in a power series in time including terms $\omega_{21}$.
It is easily shown that the first $\omega_{21}$ term corresponds to the second derivative, $\ddot{a_2}(t)$. The difference between the exact and the analytic solutions for the transition amplitudes is $ \Delta a_2(t) \approx \frac{1}{2}\omega_{21}\chi t^2 $, and
\begin{equation}\label{delta P}
\Delta P(t) = |\Delta a_2(t)|^2 \approx \frac{1}{4}\omega_{21}^2 \chi^2 t^4 \ .
\end{equation}
By the time the occupation probability $P_2$ reaches its first maximum at $t_0=T/4=\pi/2 \omega$, the difference becomes
\begin{equation}\label{delta P 2}
\Delta P(t_0) \approx \frac{1}{4}(\frac{\pi}{2})^6(\frac{\omega_{21}}{\omega})^2 \ .
\end{equation}
Comparison of Eqs. (\ref{delta P 2}) and (\ref{estim1}) shows that the choice of the field frequency, $\omega$, involves a trade-off between the duration of the state and the population leakage.
The higher the frequency, $\omega$, the smaller the population leakage, $\Delta P$.
But the higher $\omega$, the shorter the duration $T_s$ of the population transfer.

\subsection{Calculations of $2s-2p$ transitions in hydrogen}
\begin{figure}
\scalebox{0.7}{\includegraphics{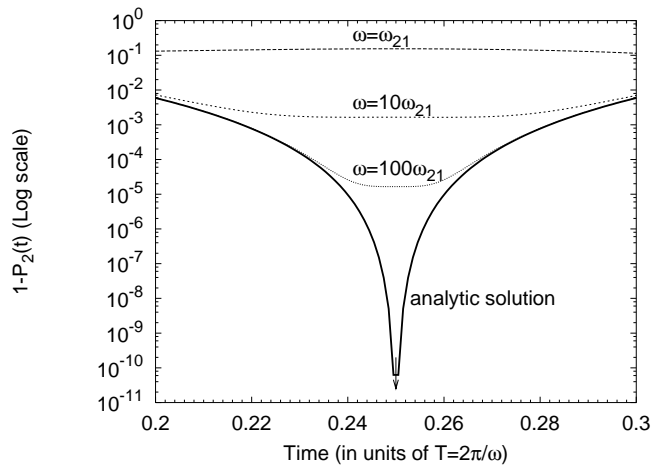}}
\caption{Occupation probability $P_2(t)$ in the vicinity of $t_0=T/4$: analytical solution, Eq.(\ref{f9}), and numerical calculations for different values of the ratio $\omega/\omega_{21}$ (1, 10 and 100). Deviation from the analytic solution decreases as $(\omega_{21}/\omega)^2$ (cf. Eqn. (\ref{delta P 2}))} 
\label{fig3}
\end{figure}

Here we compare the full (numerical) calculations and the analytical solution for a $2s-2p$ transition in hydrogen.
The full calculations were done by numerically integrating Eqs. (\ref{f6}) using a standard fourth order Runge-Kutta algorithm.
In hydrogen the energy separation of $2s$ and $2p$ levels (Lamb shift) is $4.37*10^{-6}eV$, while the next available level ($3p$) is $\approx 1.89 eV$ away. 
Therefore, one can choose the frequency of the external field $\omega$ such that 
$\omega_{21} \ll \omega \ll \omega_{2s3p}$, i.e. both the degenerate level approximation (limit $\omega_{21} \rightarrow 0$) and the two-state model can be used.

As shown in figure \ref{fig3}, the difference between the full and analytic solutions for $P_1(t)=1-P_2(t)$ in the vicinity of $t_0=\pi/2\omega$  is large when $\omega=\omega_{21}$. 
But it decreases for $\omega > \omega_{21}$. 
The difference is less than 1\% for $\omega/\omega_{21}=10$, and  0.01\% for $\omega/\omega_{21}=100$. 
Since that difference is proportional to  $(\omega_{21}/\omega)^2$, one can go up in frequency as high as fractions of $eV$ (that brings the difference between the approximate and exact solution down to $10^{-10}-10^{-11}$) and yet be far sensibly from the nearest available ($2s-3p$) resonant transition  frequency.
Therefore, a  radiation field with the wavelength from a few $\mu m$ (and the intensity of the order $10^{12} W/cm^2$) up to a few $cm$ (and the intensity of the order $10^4 W/cm^2$) can be used for $2s-2p$ transitions in hydrogen. 
The probability of the multiphoton excitation to the $3p$ state (or any other state, including continuum) also seems to be small for the range of the external field frequencies under consideration. 

In summary, our analytic approximation appears to be valid for $2s-2p$ transitions in hydrogen over a broad range of field frequencies.

\subsection{Changing the shape of $V(t)$}\label{shape}
We have noted previously that the choice of the field frequency is a trade-off between two competing factors: reducing the population leakage and increasing the duration of the populated  state. 
If wishes to obtain a long lasting populated state with very small leakage, it may be that 
both requirements for the system cannot be met simultaneously for a single frequency radiation field. 
In this case it may be possible to use another form of the interaction potential. 
In particular, one may change the shape of the "flat" part of the probability (cf. figure \ref{fig1}) by using different shapes for the external potential $V_{ext}(t)$ . For an arbitrary external potential the formulas for the occupation probabilities can be written in the form:
\begin{eqnarray}\label{a4}
&& P_1(t) = \cos^2 \lbrack \int_0^t V_{21}(\tau)d\tau \rbrack \nonumber\\
&& P_2(t) = \sin^2 \lbrack \int_0^t V_{21}(\tau)d\tau \rbrack \ .
\end{eqnarray}

A Taylor series expansion can be used for choosing the shape of the external potential. For example, one can use the potential with the first non-vanishing derivative of the order higher than four to make the shape of the populated state even flatter. 
The derivatives can be calculated using the general formula for the n$th$ derivative of a composite function \cite{Gradshteyn}, which in this case takes the form 
\begin{equation}\label{gradshteyn}
\frac{d^n P_2}{d t^n}= \sum \frac{n!}{i! j! \dots k!} \frac{d^m F}{d y^m}
(\frac{y^{\prime}}{1!})^i (\frac{y^{\prime \prime}}{2!})^j \dots (\frac{y^{(l)}}{l!})^k \ ,
\end{equation}
where $P_2(t)=F(y)=\sin^2 y \ $, $ \ y=y(t)=\int_0^t V_{21}(\tau) \ d\tau \ $, $ \ \sum$ indicates summation over all solutions in non-negative integers of the equation $i+2j+\dots +lk=n$ and $m=i+j+ \dots +k$.

In principle, Eqn.(\ref{gradshteyn}) tells us how to shape $V(t)$ to control the population transfer.
As one can see from Eq.(\ref{gradshteyn}), one may eliminate all terms up to the order $\tau^k$ by choosing the potential for which all derivatives up to the order $(k-1)$ are zero at the point $t=t_0$. 
That will "flatten" the shape of the populated (or depopulated) state, i.e. allow one to use smaller frequency (and, as a result, an increased duration of the state) to achieve the same level of population leakage. 
Therefore, the shape of an external potential can be used (along with the choice of the field frequency) for quantum control.
For example, in a truly two-level system, choosing the potential of the form
$$V(t)=\frac{\pi}{2}\delta(t-t_0)$$
leads to 
\begin{eqnarray}\label{f10}
&& P_1(t)=1-\Theta(t-t_0) \nonumber\\
&& P_2(t)=\Theta(t-t_0) \ .
\end{eqnarray}
This represents complete and immediate population inversion at $t=t_0$.
However, Fourier transformation of $V(t)$ now contains all frequencies, so one can no longer use a two-state approximation for the $2s-2p$ transition in hydrogen, since higher states become necessarily involved.  

\section{Discussion}\label{Sec:Disc}
In this paper we have used the $2s-2p$ transition in hydrogen as an example of population transfer. 
The same approach can be used for any other atomic or molecular system that has a similar pattern of energy levels (two levels located close one to another and far from the other levels). 
As we have shown in Eq.(\ref{delta P 2}), for an external field with a single frequency $\omega$, the difference between the exact calculations and the analytic approximation varies as $(\omega_{21}/\omega)^2$ for $\omega_{21}/\omega \ll 1$, and does not depend on the internal structure of the atom or molecule.
This feature opens the possibility for using different systems with different values of the transition frequency, $\omega_{21}$ and different ranges for the field frequency, $\omega$.

A downside of analyzing more complex atomic or molecular systems is that analytic expressions for the orbital functions (and, consequently, for the matrix element $V_{21}$ in Eq. (\ref{a4}))
are not always available.
Then genetic algorithms (GA) may be used to choose a shape for the external potential, $V(t)$.
The application of GA with active feedback to the selective breaking and making chemical bonds in polyatomic molecules has been discussed in detail by Rabitz {\it et al } \cite{Rabitz92, Rabitz93}, and tested experimentally \cite{Levis99, Levis01}).
As in any optimization scheme, the effectiveness of the GA increases significantly when the initial value of the parameter close to the optimal one is chosen.
One might combine our method with GA using approximate analytic orbital functions to choose the "starting" form of the potential, estimate $T_s$ and $P_{cr}$, and then employ the GA scheme.

For $2s-2p$ transitions in hydrogen, sources of the microwave radiation with corresponding intensities may now be available, which can be used to test our model.
Metastable excited $H(2s)$ atoms could interact with a low frequency (e.g. microwave) radiation.
Since the lifetime of the $2p$ state is much shorter than the $2s$ state, the population of the $2p$ state may be monitored by observing photons emitted in $2p-1s$ transitions.
In these experiments the smallness of the Lamb shift requires the use of the temperatures below the $1 \ mK$  to exclude the thermal transitions between the levels.

Finally we note that the approximation introduced here is related to the removal of time ordering in the time evolution operator \cite{mgbook}.  
The full evolution operator may be expressed as 
$U(t,t_0)= T \exp(-i \int _{t_o}^t V(t') dt')$, where $T$ is the Dyson time ordering operator
that puts the sequence $V(t_n) ... V(t_2) V(t_1)$ in the order of increasing time.  
When time ordering is removed $T \to 1$ and $U(t,t_0 )\to   \exp(-i \int _{t_o}^t V(t') dt')$.
In a two state system this corresponds to $a_1 = \cos(\int _{t_o}^t V(t') dt')$ and
$a_2 = i \sin(\int _{t_o}^t V(t') dt')$, as used in this paper.
This approximation has been recently studied in the context of second order perturbation
theory \cite{mg01,gm01,gm02}, where it has been shown that the time evolution of different
electrons becomes uncorrelated as $T \to 1$.
Corrections to this Magnus-like approximation \cite{Magnus} lead to correlated propagation that is non local in time.  These corrections correspond to quantum fluctuations in the energy in short lived intermediate states arising from coupling of a quantum system with its macroscopic environment \cite{mg02}.

\section{Summary}

A new analytical solution of two coupled channel equations has been found that enables one to temporally control the electron population of $2s$ and $2p$ states in hydrogen by using a time varying external field.
In addition, the population leakage can both be easily estimated, and than be further controlled by changing the shape of the external field.
Our method can be applied to various systems. 

\begin{acknowledgments}
We thank J.H. Eberly who encouraged us to use non-perturbative methods, showed us how coupled channel equations for photon interactions work and provided insightful examples.
This work was supported by the Division of Chemical Sciences, Office of Science, U.S. Department of Energy.
\end{acknowledgments}

\end{document}